\documentclass[a4paper]{article}

\usepackage{epsfig}
\usepackage{natbib}

\begin{document}

\title{Chaotic hopping between attractors in neural networks}
\author{J. Marro, J.J. Torres\footnote{Corresponding author: Joaquin J. Torres, Department of Electromagnetism and Physics of the Matter, University of Granada, E-18071 Granada Spain; Tlf: (+34) 958244014; Fax: (+34) 95824 23 53 \hbox{email: jtorres@onsager.ugr.es}}, and J.M. Cortes\footnote{Present address: Institute for Adaptive and Neural Computation, School of Informatics, The University of Edinburgh, 5 Forrest Hill, Edinburgh EH1 2QL, Scotland UK}\\
\small Institute \textit{Carlos I} for Theoretical and Computational
Physics, and \\
\small Departamento de Electromagnetismo y F\'{\i}sica de la Materia,\\
\small University of Granada, E-18071--Granada, Spain.}
\date{}
\maketitle

\begin{abstract}
We present a neurobiologically--inspired stochastic cellular automaton whose
state jumps with time between the attractors corresponding to a series of
stored patterns. The jumping varies from regular to chaotic as the model
parameters are modified. The resulting irregular behavior, which mimics the 
\textit{state of attention} in which a systems shows a great adaptability to
changing stimulus, is a consequence in the model of short--time presynaptic
noise which induces \textit{synaptic depression}. We discuss results from
both a mean--field analysis and Monte Carlo simulations.

\end{abstract}

\section{Introduction}

Analysis of electroencephalogram time series, though perhaps not conclusive
yet, suggest that some of the brain high level tasks might require chaotic
activity and itinerancy \citep{barrie,tsuda,korn,freeman3}. As a
matter of fact, following the observation of constructive chaos in many
natural systems \citep{ca96,ca01,ca03}, it has been reported some evidence
that chaos may, for example, enhance sensitivity by inducing a critical
state of synchronization during expectation and attention \citep%
{hansel2}, and perhaps provide an efficient means to discriminate
different (e.g.) olfactory stimuli \citep{ashwin05}. Consequently,
there has recently been some effort in incorporating constructive chaos in neural
network modeling \citep{wang,bolle,domin2,caro2,poon,mai,kata}.
Concluding on the significance of chaos in neurobiological systems is still
an open issue \citep{rabi,faure,korn}, however.

As a new effort towards better understanding this problem, in the present
paper we present, and study both analytically and numerically, a \textit{%
neural automaton} which exhibits chaotic behavior. More specifically, it
shows sort of \textit{dynamic} associative memory, consisting of chaotic
hopping between the stored memories, which mimics the brain states of
attention and searching. The model is a neurobiologically inspired cellular
automaton, in which dynamics concerns the whole, which is
simultaneously updated ---instead of sequentially updating a small
neighborhood at each time step. This automaton (or \textit{Little dynamics}) 
strategy has already revealed efficient in modeling several
aspects of associative memory \citep{gang,cortes}. Interesting enough,
concerning this property, neural automata often exhibit more interesting
behavior than their Hopfield--like, sequentially--updated counterparts, in
spite of the fact that any two successive states are stronger correlated in
the sequential case. Therefore, we extend here to cellular automata our
recent study of the effects of synaptic \textquotedblleft
noise\textquotedblright\ on the stability of attractors in Hopfield--like
networks \citep{cortes2}. We demonstrate that, in our automaton, a certain
type of synaptic fluctuations determine an interesting retrieval process.
The model synaptic fluctuations are coupled to the presynaptic activity in
such a way that \textit{synaptic depression} occurs. This phenomenon, which
has been observed in actual systems, consists in a
lowering of the neurotransmitters release after a period of intense
presynaptic activity \citep{tsod,pantic}. Our model fluctuations happen to destabilize the
memory attractors and are shown to induce, eventually, regular and even
chaotic dynamics between the stored patterns. Confirming expectations
mentioned above, we also show that our model behavior implies a high
adaptability to a changing environment, which seems to be one of the nature
strategies for efficient computation \citep{lu,schwei}.

\section{The model}

Let a set of $N$ binary neurons, $\mathbf{S=}\{s_{i}=\pm 1;i=1,\ldots ,N\},$
connected by synapses of intensity:%
\begin{equation}
w_{ij}=w_{ij}^{\mathrm{L}}x_{j},\,\,\,\,\forall i,j.  \label{w}
\end{equation}%
Here, $x_{j}$ stands for a random variable, and $w_{ij}^{\mathrm{L}}$ is an
average weight. The specific choice for the latter is not essential here
but, for simplicity and reference purposes, we shall consider a Hebbian 
\textit{learning rule} \citep{amitB}. That is, we shall assume in the
following that synapses \textit{store }a set of $M$ binary patterns, $%
\mathbf{\xi }^{\mu }=\left\{ \xi _{i}^{\mu }=\pm 1;i=1,\ldots ,N\right\} ,$ $%
\mu =1,...,M,$ according to the prescription, $w_{ij}^{\mathrm{L}%
}=M^{-1}\sum_{\mu }\xi _{i}^{\mu }\xi _{j}^{\mu }.$

The set $\mathbf{X=}\{x_{j}\}$ of random variables is intended to model some
reported fluctuations of the synaptic weights. To be more
specific, the multiplicative noise in (\ref{w}), which was recently used to
implement a variation of the Hopfield model \citep{cortes2}, may have
different competing causes in practice, ranging from short--length
stochasticities, e.g., those associated with the opening and closing of the
vesicles and with local variations in the neurotransmitters concentration,
to time lags in the incoming long--length signals \citep{franks}. These
effects will result in short--time, i.e., relatively fast microscopic noise.
As a matter of fact, the typical synaptic variability is reported to occur
on a time scale which is small compared with the characteristic system
relaxation \citep{zadorJN}. Therefore, as far as $\mathbf{X}$ corresponds to
microscopic fast noise, neurons will evolve as in presence of a steady
distribution, say $P^{\mathrm{st}}(\mathbf{X}|\mathbf{S}).$ It follows that
such noise will modify the \textit{local fields}, $h_{i}(\mathbf{S},\mathbf{X%
})=\sum_{j\neq i}w_{ij}x_{j}s_{j},$ i.e., the total presynaptic current
which arrives to the postsynaptic neuron $s_{i},$ which one may assume to be
given in practice by%
\begin{equation}
\overline{h_{i}}(\mathbf{S})=\int_{\mathbf{X}}h_{i}(\mathbf{S},\mathbf{X})P^{%
\mathrm{st}}(\mathbf{X}|\mathbf{S})\mathbf{\mathrm{d}}\mathbf{X}.
\label{heff}
\end{equation}%
This, which is a main feature of our automaton, amounts to assume that each
neuron endures an effective field which is, in fact, the average
contribution of all possible different realizations of the actual field \citep%
{bibi}. This situation has been formally discussed in detail in Refs.\citep%
{torres2,marro}. It may be noticed that, consistently with the choice
of a binary, $\pm 1$ code for the neurons activity, we are assuming zero
thresholds, $\theta _{i}=0,$ $\forall i,$ in the following; this is relevant
when comparing this work with some related one, as discussed below.

Next, one needs to model the noise steady distribution. Motivated by some
recent neurobiological findings, we would like this to mimic short--term 
\textit{synaptic depression} \citep{tsod,pantic}. This refers to the
observation that the synaptic weight decreases under repeated presynaptic
activation. The question is how such mechanism may affect the automata (and,
in turn, actual systems) dynamics. For simplicity, we shall assume
factorization of the noise distribution, i.e., we assume the simplest case $%
P^{\mathrm{st}}(\mathbf{X}|\mathbf{S})=\prod_{j}P(x_{j}|\mathbf{S}),$ and 
\begin{equation}
P(x_{j}|\mathbf{S})=\zeta \left( \vec{m}\right) \mathrm{\ }\delta
(x_{j}+\Phi )+\left[ 1-\zeta \left( \vec{m}\right) \right] \mathrm{\ \ }%
\delta (x_{j}-1).  \label{bimod}
\end{equation}%
Here, $\vec{m}=\vec{m}(\mathbf{S})$ stands for the overlap vector of
components $m^{\mu }(\mathbf{S})=N^{-1}\sum_{i}\xi _{i}^{\mu }s_{i},$ and $%
\zeta \left( \vec{m}\right) $ is an increasing function of $\vec{m}$ to be
determined. The choice (\ref{bimod}) amounts to assume that, with
probability $\zeta \left( \vec{m}\right) ,$ i.e., more likely the larger $%
\vec{m}$ is, which implies a larger net current arriving to the postsynaptic
neurons, the synaptic weight will be depressed by a factor $-\Phi .$
Otherwise, the weight is given the chosen average value, see equation (\ref%
{w}). Interesting enough, (\ref{bimod}) clearly induces some non--trivial
correlations between synaptic noise and neural activity. This is an
additional bonus of our choice, as it conforms the general expectation that
processing of information in a network will depend on the noise affecting
the communication lines and vice versa \citep{cortes2}. Looking for an
increasing function of the total presynaptic current with proper
normalization, a simple choice for the probability in (\ref{bimod}) is $%
\zeta \left( \vec{m}\right) =\left( 1+\alpha \right) ^{-1}\sum_{\mu }\left[
m^{\mu }\left( \mathbf{S}\right) \right] ^{2},$ where $\alpha =M/N$ is the
load parameter or network capacity. It then follows after some simple
algebra that the resulting fields are%
\begin{equation}
\overline{h_{i}}(\mathbf{S})=\left\{ 1-\gamma \sum_{\mu }\left[ m^{\mu
}\left( \mathbf{S}\right) \right] ^{2}\right\} \sum_{\nu }\xi _{i}^{\nu
}m^{\nu }\left( \mathbf{S}\right) ,  \label{lfaprox}
\end{equation}%
where $\gamma \equiv \left( 1+\Phi \right) \left( 1+\alpha \right) ^{-1}.$
Notice that this precisely reduces for $\Phi \rightarrow -1$ to the local
fields in the Hopfield model in which the synaptic weights do not fluctuate
but are constant in time \citep{amitB}.

Time evolution is due to competition between these fields, which contain the
effects of synaptic \textit{noise,} and some additional natural
stochasticity of the neural activity. In accordance with a familiar
hypothesis, we shall assume this stochasticity controlled by a
\textquotedblleft temperature\textquotedblright\ parameter, $T,$ which
characterizes an underlying \textquotedblleft thermal
bath\textquotedblright\ \citep{marro}. Consequently, evolution is by the
stochastic equation $\Pi _{t+1}(\mathbf{S})=\sum_{\mathbf{S^{\prime }}}\Pi
_{t}(\mathbf{S^{\prime }})\Omega (\mathbf{S^{\prime }}\rightarrow \mathbf{S}%
),$ where the probability per unit time of a transition is%
\begin{equation}
\Omega (\mathbf{S^{\prime }}\rightarrow \mathbf{S})=\prod_{i=1}^{N}\omega
(s_{i}^{\prime }\rightarrow s_{i}).  \label{jp}
\end{equation}%
For simplicity and concreteness, we take $\omega (s_{i}^{\prime }\rightarrow
s_{i})\propto \Psi \left[ \beta _{i}\left( s_{i}^{\prime }-s_{i}\right) %
\right] ,$ where $\beta _{i}\equiv T^{-1}\overline{h_{i}}(\mathbf{S^{\prime }%
}),$ and $\overline{h_{i}}(\mathbf{S^{\prime }})$ independent of $%
s_{i}^{\prime }$, which is a good approximation for a sufficiently large
network (technically, this is an exact property in the \textit{thermodynamic
limit} $N\rightarrow \infty $). The function $\Psi $ is arbitrary except
that, in order to obtain well defined limits, we require that $\Psi (u)=\Psi
(-u)\exp (u),$ $\Psi (0)=1$ and $\Psi (\infty )=0,$ which holds for a
normalized exponential function \citep{marro}. Then, consistent with the
condition $\sum_{\mathbf{S}}\Omega (\mathbf{S^{\prime }}\rightarrow \mathbf{S%
})=1,$ we take%
\begin{equation}
\omega (s_{i}^{\prime }\rightarrow s_{i})=\Psi \lbrack \beta
_{i}(s_{i}^{\prime }-s_{i})]\left[ 1+\Psi \left( 2\beta _{i}s_{i}^{\prime
}\right) \right] ^{-1}.  \label{tnorm}
\end{equation}

\section{Main results}

It is obvious that the above may be adapted to cover other, more involved
cases \citep{cortes2}, but this is enough to our purposes here. In fact, Monte
Carlo simulations reveal some new interesting facts as compared with the
case of sequential updating \citep{cortes2}. To begin with, figure \ref%
{figure1} illustrates a much varied landscape, namely, the occurrence of
fixed points, cycles, regular and irregular hopping between the attractors.
This may also be obtained analytically in the mean--field approximation $%
s_{i}=\langle s_{i}\rangle $ $\forall i$ \citep{amitB}. We then obtain
for $M=1$ a discrete map which describes time evolution of the overlap $%
m\equiv m^{1}$ as%
\begin{equation}
m_{t+1}=\tanh \{T^{-1}m_{t}[1-m_{t}^{2}(1+\Phi )]\}.  \label{mfdeo1p}
\end{equation}%
As one varies here the \textquotedblleft temperature\textquotedblright\ $T$
and the depressing parameter $\Phi ,$ it follows a varying situation in
perfect agreement with the Monte Carlo simulations, as one should have
expected for a fully connected network. In particular, figure \ref{figure2}
shows the occurrence of chaos in a case in which thermal fluctuations are
small compared to the synaptic noise. That is, the Lyapunov exponent, $%
\lambda ,$ corresponding to the dynamic mean--field map shows different
chaotic windows, i.e., $\lambda >0,$ as one varies $\Phi $ for a fixed $T.$
As illustrated also in figure \ref{figure2}, dynamics is stable for $\Phi =-1
$, i.e., in the absence of any synaptic noise, and the only solutions then
correspond to the ones that characterize the familiar Hopfield case with
parallel updating. As $\Phi $ is increased, however, the system
tends to become unstable, and transitions between $m=1$ and $m=-1$ then
eventually occur that are fully chaotic.

There is also chaotic hopping between the attractors when the system stores
several patterns, i.e., for $M>1.$ In this case, we obtain the more complex, multidimensional map:
\begin{equation}
m^{\nu}(\mathbf{S})(t+1)=\frac{1}{N}\sum_i \xi^\nu_i \tanh [\beta \overline{h_{i}}(\mathbf{S})(t) ] \quad \forall \nu=1,\ldots M.
\end{equation}
This is to be numerically iterated. The simplest order parameter to monitor
this is:%
\begin{equation}
\zeta =\frac{1}{1+\alpha }\sum_{\mu }\left( m^{\mu } (\mathbf{S})\right) ^{2}.
\end{equation}%
This is shown in figure \ref{figure3} as a function of $\Phi .$ The graph
clearly illustrates a region of irregular behavior which has a width $\Delta
\Phi _{c}$ defined as the distance, in terms of $\Phi ,$ from the first
bifurcation to the last one. Interesting enough, we find that the width of
this region is practically independent of the number of patterns; that is,
we find that $\Delta \Phi _{c}=0.575\pm 0.005$ as $M$ is varied within the
range $M\in \left[ 1,50\right] .$ This suggests that the chaotic behavior
which occurs for depressing fast synaptic fluctuations, i.e., for any $\Phi
>-1,$ does not critically depend on the automaton capacity but the model
properties are rather robust and perhaps independent of the number of stored
patterns within a wide range. One may expect, however, that some of the
interesting model properties will tend to wash out as the load parameter
increases macroscopically, i.e., as $M\rightarrow \infty .$

\section{Discussion and further results}

Motivated by the fact that analysis of brain waves provides some indication
that the chaos--theory concept of \textit{strange attractor} may be relevant
to describe some of the neural activity, we presented here a
neurobiologically--inspired model which exhibits chaotic behavior. The model
is a (microscopic) cellular automaton with only two parameters, $T$ and $%
\Phi ,$ which control the thermal stochasticity of the neural activity and
the depressing effect of (coupled) fast synaptic fluctuations, respectively.
Our system reduces to the Hopfield case with Little dynamics (parallel
updating) only for $\Phi =-1.$

Our main result is that, as described in detail in the previous section, the
automaton eventually exhibits chaotic behavior for $\Phi \neq -1,$ but not
for $\Phi =-1,$ nor in the case of sequential, single--neuron updating
irrespective of the value of $\Phi $ \citep{cortes2}. It also follows from
our analysis above that further study of this system and related automata is
needed in order to determine other conditions for chaotic hopping. For
example, one would like to know if synchronization of all variables is
required, and the precise mechanism for moving from regular to irregular
behavior as $\Phi $ is slightly modified. We are pursuing the present effort
along this line \citep{cortes2}, and present some related preliminary
conclusions below.

This is not the first time in which chaos is reported to occur during the
retrieval process in attractor neural networks; see, for instance, \citep%
{wang,bolle,domin2,poon,caro2,mai,kata}. One may say, however,
that we provide in this paper a more general and microscopic setting than
before and, in fact, the onset of chaos here could not be phenomenologically
predicted. That is, the same microscopic mechanism, namely (\ref{w}) and (%
\ref{bimod}), does not imply chaotic behavior if updating is by a sequential
single--variable process \citep{cortes2}. Another possible comparison is by
noticing that, in any case, whether one proceeds more or less
phenomenologically, the result is a map $m_{t+1}=G(m_{t}).$ We obtained the
gain function $G$ after coarse graining of (\ref{lfaprox})--(\ref{tnorm}),
and the Monte Carlo simulations fitting the map behavior just involve
neurons subject to the local fields (\ref{heff}), so that we are only left
in the two cases with the noise parameter $\Phi $ to be tuned. In contrast,
some related works
, in order to deepen more directly on the possible origin of chaos, use the
gain function itself as a parameter. It is also remarkable that, e.g., in 
\citep{domin2} and some related work \citep{caro2,mai,kata}, the gain
function is phenomenologically controlled by tuning the neuron threshold for
firing, $\theta _{i}.$ The threshold function thus becomes a relevant
parameter, and it ensues that any meaningful chaos in this context requires
non--zero threshold. This is because, in these cases, the local fields and,
consequently, the overlaps, are lineal, which forces one to induce chaos by
other means. Interesting enough, our gain function in (\ref{mfdeo1p}) has
either a sigmoid shape or an oscillating one, as illustrated for $T=0$ in
figure \ref{figure4}. Only the latter case allows for hopping between the
attractors and, eventually, for chaotic behavior.

Finally, we demonstrate an interesting property of our automaton during
retrieval. This is the fact that, in the chaotic regime, the system is
extremely susceptible to external influences. A rather stringent test of
this is its behavior concerning mixture or \textit{spin--glass} steady
states, which are unsuited in relation with associative memory. Even though
these states may occur at low $T,$ this system ---unlike other cases---
easily escapes from them under a very small external stimulus. This is
illustrated in figure \ref{figure5} which also demonstrates a general
feature, namely, some strong correlation between chaos and a vivid response
to the environment. This nicely conforms expectations mentioned above, in
the introduction of this paper, that chaotic itinerancy might be a rather
general strategy of nature.

\section{Acknowledgments}

We acknowledge with thanks very useful discussions with David R.C. Dom\'{\i}%
nguez, Pedro L. Garrido, Sabine Hilfiker and Hilbert J. Kappen, and
financial support from MEyC and FEDER, project No. FIS2005-00791.


\newpage
\section*{Figure Captions}

{\bf Figure 1:} Monte Carlo time--evolution of the overlap between the automaton
current state and the given stored pattern for $M=1,$ $N=10$$^{\mbox{4}}$
neurons, $T=0.1,$ and different values of $\Phi ,$ as indicated. This
illustrates, from top to bottom, the fixed point solution in the absence of
any synaptic noise, i.e., $\Phi =-1$, a cyclic behavior, the onset of
irregular periodic behavior, and fully irregular and regular jumping between
the two attractors corresponding, respectively, to the given pattern, $%
m\equiv m^{1}=1,$ and its \textit{anti--pattern} $m=-1$ ---the only
possibilities in this case with $M=1.$
\\\\
\noindent
{\bf Figure 2:} Bifurcation diagram and associated Lyapunov exponent demonstrating
chaotic activity for some (but not all) values of the depressing coefficient 
$\Phi .$ The upper graph shows, for $M=1,$ the steady overlap between the
current state and the given pattern as a function of $\Phi .$ This is from
Monte Carlo simulations of a network with $N=10^{4}$ neurons. The bottom
graph depicts the corresponding Lyapunov exponent, $\protect\lambda $, as
obtained from the map (\protect\ref{mfdeo1p}). This confirms the existence
of \textit{chaotic windows,} in which $\protect\lambda >0$. The \textit{%
temperature} parameter is set $T=0.1$ in both cases; this is low enough so
that the effect of thermal fluctuations is negligible compared to that of
synaptic noise.
\\\\
\noindent
{\bf Figure 3:} The function $\protect\zeta \left( \Phi \right) ,$ {as defined in
the main text, obtained from Monte Carlo simulations at }$T=0.15${\ for }$%
N=10^{4}${\ neurons and }$M=20${\ stored patterns generated at random. A
region of irregular behavior which extends for }$\Delta \Phi _{c},${\ as
indicated, is depicted. The insets show the time evolution of four out of
the 20 overlaps within the irregular region, namely, for }$\Phi =0.11.$
\\\\
\noindent
{\bf Figure 4:} The gain function in (\protect\ref{mfdeo1p}) versus $m_{t}$ for $T=0
$ and different values of $\Phi $, as indicated. It is to be remarked that
this function is non-sigmoidal, namely, oscillatory, which allows for
hopping between the attractors for $\Phi >0,$ while it is monotonic in the
Hopfield case $\Phi =-1.$
\\\\
\noindent
{\bf Figure 5:} Time evolution of the overlap $m^{\protect\mu }$ in
a Monte Carlo simulation with $N=$10$^{4}$ neurons, $M=4$ stored (random)
patterns, $T=0.05,$ and, from top to bottom, $\Phi =-0.2,$ $-0.1,$ $0.12,$
and $0.2.$ This illustrates that, under regular behavior (as for the first
two top graphs and the bottom one), the system is unable to respond to a
week external stimulus. This is simulated as an extra local field, $h_{i}^{ 
\mbox{\protect
\footnotesize ext}}=\protect\delta \protect\xi _{i}^{\protect%
\mu }$, where $\protect\delta =0.05$ and $\protect\mu $ changes $(\protect%
\mu =1,$ $2,$ $3,$ $4,$ $1)$ every 40 MCS as indicated by P$_{\protect\mu }$
above the top graph. The situation is qualitatively different when the
regime is chaotic, as for $\Phi =0.12$ in this figure. After some wandering
in the evolution that we show here, the system activity is trapped in a
mixture state around $t=80$ MCS. However, the external stimulus induces
jumping to the more correlated attractor, and so on. That is, chaos
importantly enhances the network sensitivity. To obtain a similar behavior
during the regular regimes, one needs to increase considerably the external
force $\protect\delta .$

\newpage
\begin{figure}[t!]
\centerline{\psfig{file=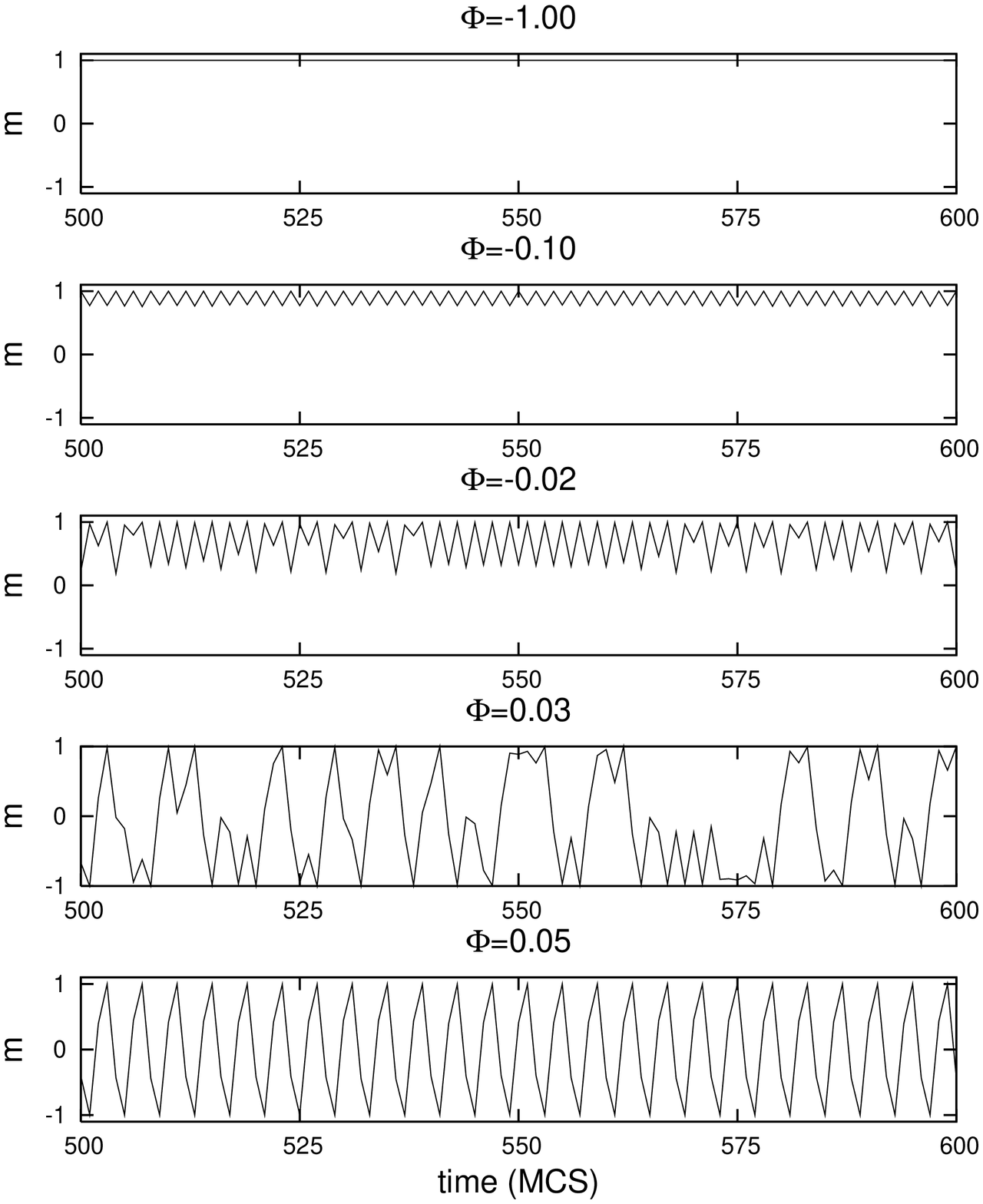,width=9.8cm}}
\caption{}
\label{figure1}
\end{figure}
\newpage

\begin{figure}[t!]
\centerline{\psfig{file=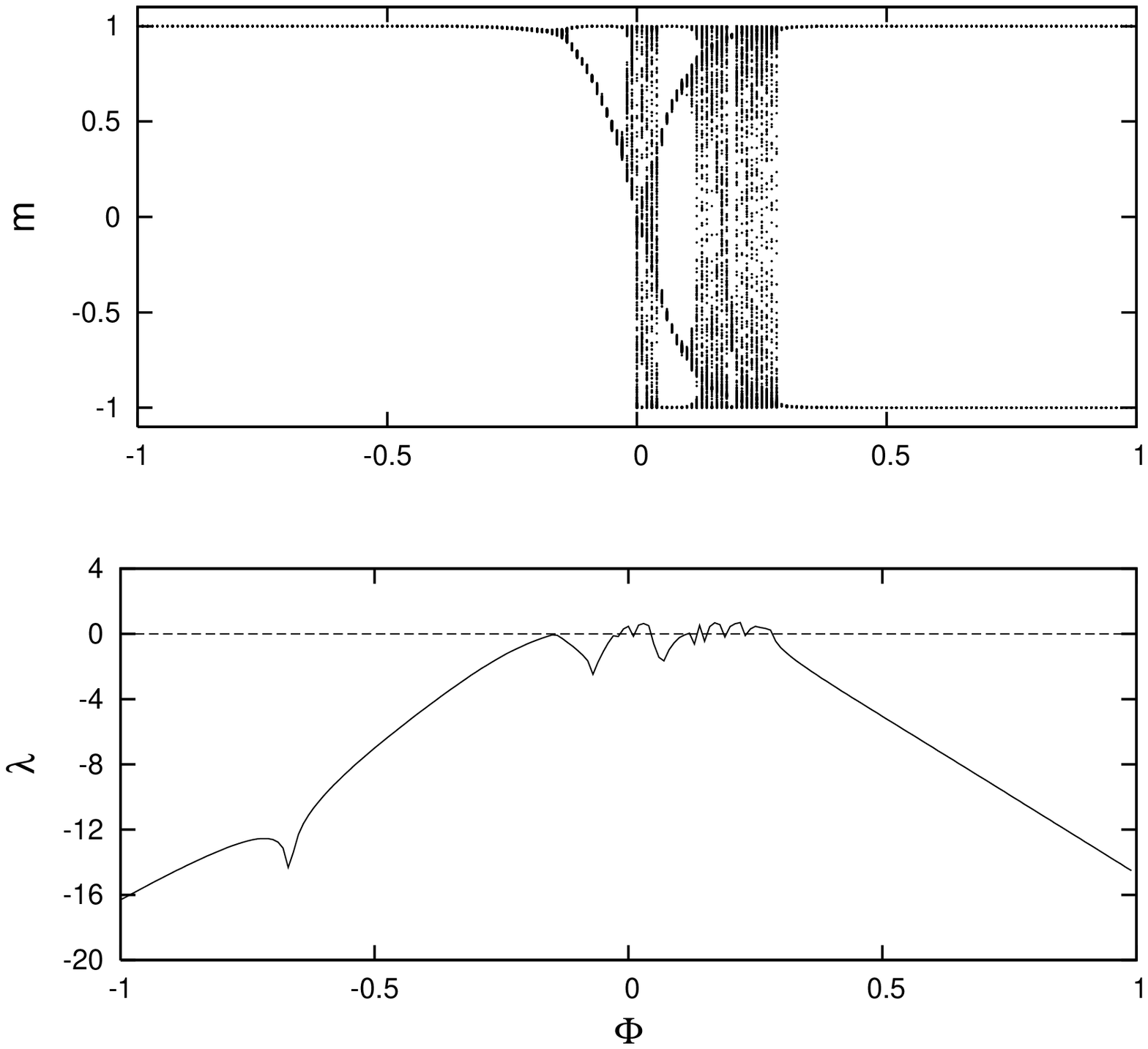,width=9.8cm}}
\caption{}
\label{figure2}
\end{figure}

\newpage

\begin{figure}[t!]
\centerline{\psfig{file=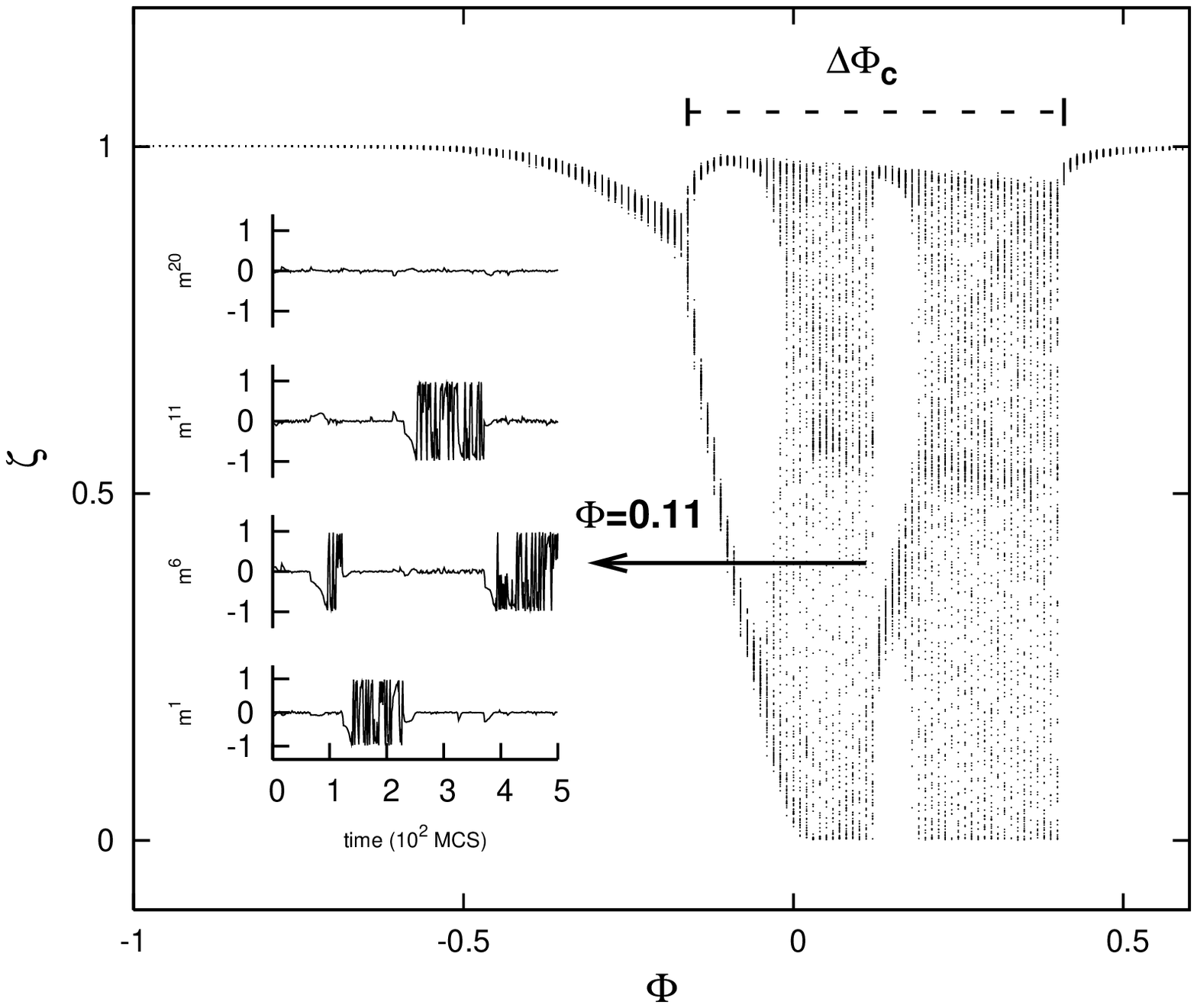,width=10.cm}} {}
\caption{}
\label{figure3}
\end{figure}

\vspace*{3cm}

\begin{figure}[ht!]
\centerline{\psfig{file=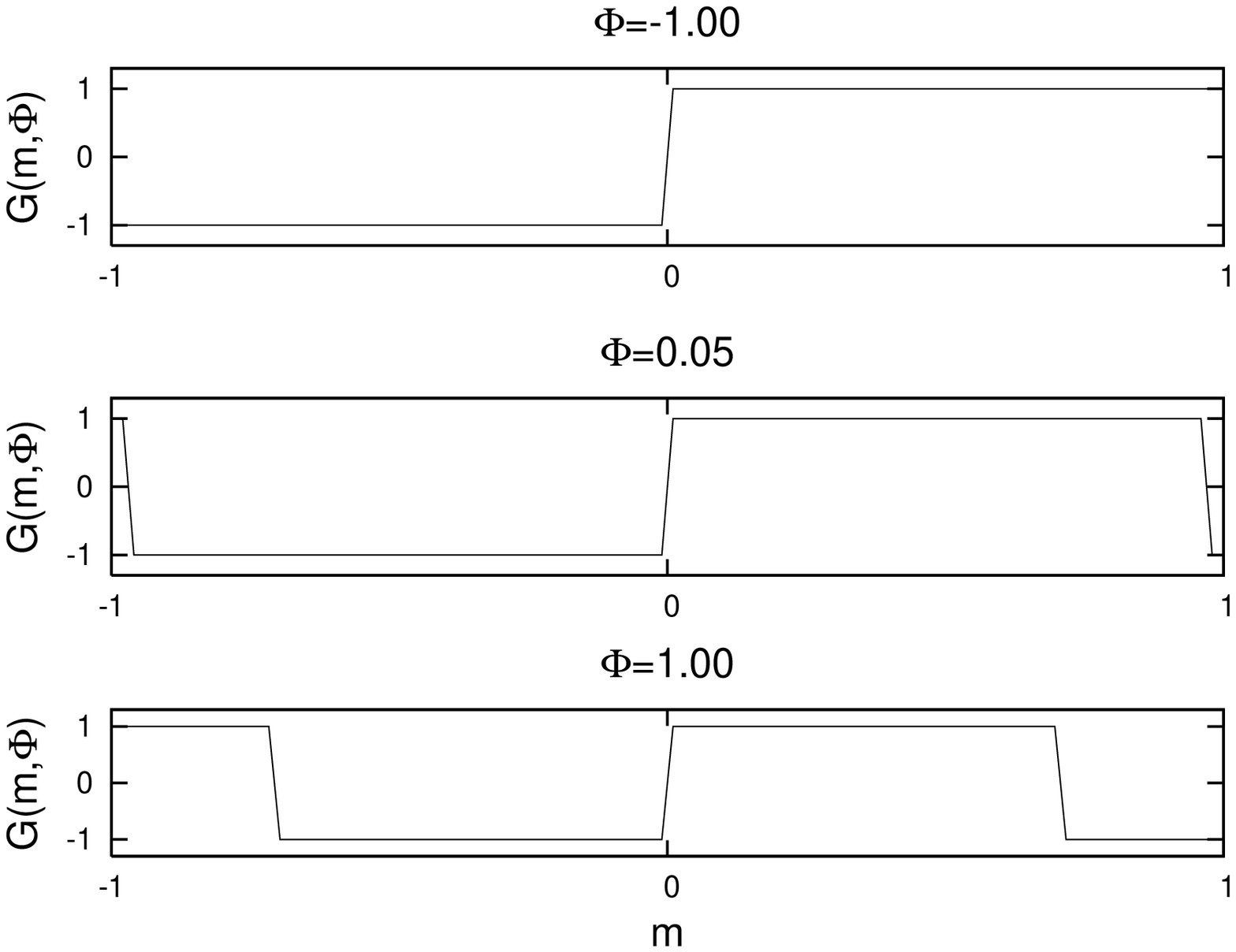,width=11.cm}}
\caption{}
\label{figure4}
\end{figure}
\vspace*{3cm}
\newpage

\begin{figure}[t!]
\centerline{\psfig{file=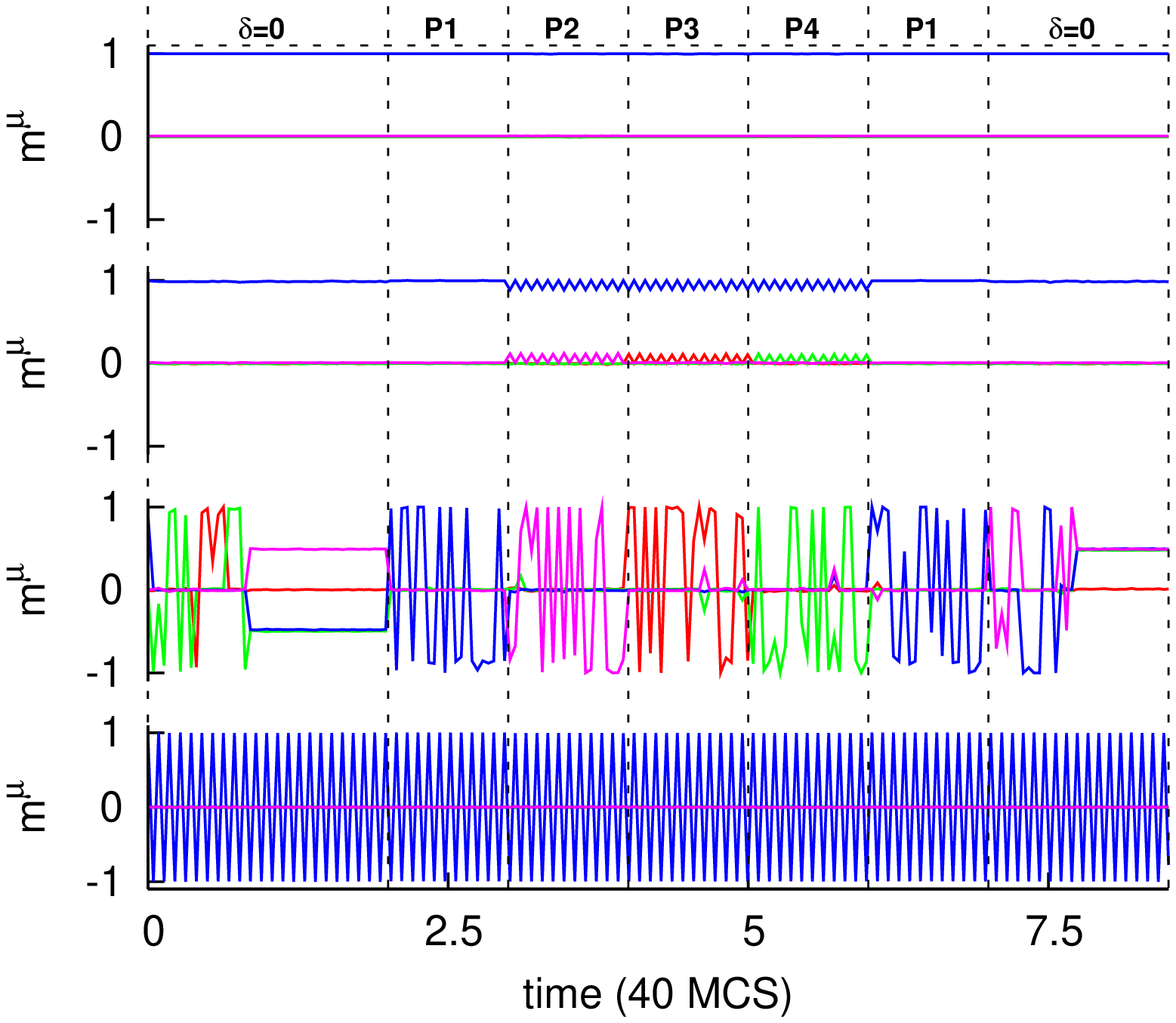,width=10.9cm}}
\caption{}
\label{figure5}
\end{figure}

\end{document}